\documentclass[manuscript,screen]{acmart}

\title[LLMs Will Change Interaction]{Large Language Models Will Change The Way Children Think About Technology And Impact Every Interaction Paradigm}

\author{Russell Beale}
\orcid{0000-0002-9395-1715}
\affiliation{
  \institution{School of Computer Science, University of Birmingham}
  \streetaddress{Edgbaston}
  \city{Birmingham}
  \postcode{B15 2TT}
  \country{UK}
}
\email{r.beale@bham.ac.uk}

\acmConference[IDC2025]{Interaction Design and Children Conference}{2025}{Reykjavík, Iceland} 

\begin{abstract}
This paper presents a hopeful perspective on the potentially dramatic impacts of Large Language Models on how we children learn and how they will expect to interact with technology.  We review the effects of LLMs on education so far, and make the case that these effects are minor compared to the upcoming changes that are occurring.  We present a small scenario and self-ethnographic study demonstrating the effects of these changes, and define five significant considerations that interactive systems designers will have to accommodate in the future.

\end{abstract}



\begin{document}
\maketitle

\section{Introduction}
This paper is a reflection on the speed with which large language models have advanced, and a prophecy on their impact on children's education and the technologies they will be using.  It is a call for education and research in this space so that we can harness this irresistible force for more good than harm, and provides some early themes for designers to consider. We firstly discuss where and how LLMs have been used in school educational settings, and then explore the new opportunities that recently released models offer.  A small-scale investigation reveals potentially large impacts on how children learn, and we highlight key things that we as a community need to be aware of.

\section{A Simple Guide to Large Language Models}
Large Language Models --- think ChatGPT, Gemini, GPT-3, CoPilot --- are  immense deep learning neural networks with exceptional numbers of parameters, which are trained to predict sequences of words, having been trained on most of the contents of the Internet.  If I asked you to complete the sentence

\begin{quote}
    Twinkle, twinkle, little star, how I wonder what you .....
\end{quote}
it is quite likely that, if you have been brought up in a Western culture, you will recognise the nursery rhyme and complete the line with
\begin{quote}
    .....are
\end{quote}
LLMs do this, but on a massive scale. As the LLM has processed much of what has ever been written, it has ingested a large number of sequences of words, and compresses them to create an internal representation. An LLM can be seen as the JPEG of the web --- it is a lossy compressed version of the internet.  What emerges from this system are responses that are natural, complex, and often insightful.  Many people are familiar with copilot being able to summarise emails or documents - long sentences are replaced  with shorter sentences it has seen in similar contexts. Where its abilities become more impressive is when you ask it to do  more complex tasks that we previously imaged required understanding and insight, such as to re-write one thing in the style of another.  When asked to help with coding, it is even more helpful --- because code is historically either right and it works, or wrong and fails, much of the code in published projects is correct, and so the LLM can provide correct code solutions for problems with remarkable abilities to refine it and improve it based on further user input.  LLMs are not new in education, but the scope of their impact may have been underestimated.

\section{Literature Review: The Evolving Landscape of LLMs in Education}
As Large Language Models (LLMs) continue to grow in sophistication and relevance, the scope of their educational applications has broadened. Early efforts to leverage LLM-based solutions for children’s learning focused primarily on tutoring, reading comprehension, and essay feedback\cite{Bommasani2021}\cite{Holmes2019}. More recent research has explored how LLMs can cultivate essential higher-order skills—such as curiosity, critical thinking, creativity, and domain-specific competencies—in settings that range from classrooms to medical training programs 
\cite{Kung2022}. In the computing domain, MacNeil et al.\cite{MacNeil2022} utilize GPT-3 to generate explanatory text for code snippets. This approach is reminiscent of earlier adaptive tutoring systems but with the added advantage of GPT-3’s ability to parse and elucidate syntactic structures in a natural-language format, underscores GPT-3’s potential to scaffold learners in debugging and conceptualizing new programming constructs.

Adaptive assessment is another promising arena for LLM deployment. Li et al.\cite{li_supporting_2024} devised a pipeline for generating diagnostic questions from textbook materials for a Biology curriculum. Their approach involved fine-tuning GPT-3 on text-based learning resources and then evaluating the model’s question quality via both automated labelling (again using GPT-3) and human expert reviews. According to their findings, many questions were deemed relevant and pedagogically sound, highlighting LLMs as a scalable solution for producing supplementary study or review materials.

Beyond text-based interactions, conversational AI in language learning has branched into multiple use-cases. A recent review by Ji et al.\cite{Ji2022} identifies five key applications of conversational AI in language education, of which the most common is deploying LLMs as chat partners,  orally or textually, to help learners practice. Jinming and Ben Kai\cite{jinming_systematic_2024} review speaking practice, while Tai \& Chen\cite{TaiChen2020} and Jeon\cite{jeon_exploring_2024}, explore how these systems can reduce foreign language anxiety and address low willingness to communicate. These studies highlight how LLM-driven conversational interfaces act as both digital tutors and emotional supports for hesitant language learners.  These experiences of improved relationships with the technology are backed up by Seymour et al. \cite{seymour_alexa_2021} who found teenagers exposed to programming Alexa found her more trustworthy and friendly afterwards, whilst Hu et al. \cite{hu_grow_2024} leveraged LLMs to improve children's mental resilience.

 LLMs are not perfect: design guidelines for children rightly emphasise privacy and transparency in data use \cite{UNICEF2025}, recognising LLMs' educational impact depends on classroom integration \cite{LeeKwon2024}.  Since they generate the most probable responses, they sometimes produce entirely fictional but plausible content, such as fabricated academic references or legal cases. Accuracy and bias remain concerns\cite{Mehrabi2021}, especially in sensitive fields like medical education\cite{Kung2022}. This highlights the need for children to critically assess LLM outputs, question their validity, and identify inconsistencies --- but is this not the aim of all education anyway?  LLMs have the potential to enhance learning universally: they adapt to different skill levels, support multiple languages, and offer both text and audio interactions, making them versatile educational tools for all learners.   

\section{New directions}
The majority use of LLMs to date has been driven by educators exploring how they can assist their students, rather than being user-led.  This is evolving just as the use of LLMs is rising in schools: Zhu et al. \cite{zhu_embracing_2024} report a study in the US in 2024 that shows more than 70\% of students used LLMs, relatively consistent across all secondary grades, 44\% using them at least once a week, whilst Williams et al. \cite{williams_examining_2023} explore the generally positive attitudes teenage learners' have to LLMs.

\subsection{How LLMs Enable Curiosity-Driven Learning}
One of the newer and more potentially revolutionary influences is the use of LLMs to positively stimulate curiosity and encourage exploration of ideas. Mintz \cite{mintz2025} employed a LLM to automate curiosity-prompting cues and encourage children to ask deeper questions to explore history. This research is grounded in psychological and neuroscientific perspectives that link curiosity to increased engagement and better retention of information \cite{kidd_psychology_2015}. By generating dynamic, context-sensitive prompts, the LLM can detect potential knowledge gaps in the learner’s responses and provide nudges that spark an “information gap” \cite{loewenstein_psychology_1994} driving the student to seek answers. Han \& Zhenyao used generative AI to enable children to create stories \cite{Han2023} and demonstrated a positive impact on literacy skills and creativity. whilst another study \cite{xu_elinors_2022}  embedded a conversational AI character into a children's science show allowing them to interact verbally: they were actively engaged and showed improved performance in immediate science assessments.

New LLMs now feature reasoning stages, providing real-time insights into how they analyze and solve problems. This transparency helps users understand their process. Combined with their patience, ability to refine responses, and flexibility in expanding or condensing arguments, LLMs serve as ideal learning companions. They function as tireless teachers who explain concepts in multiple ways and are always available. With conversational interaction, children can develop and refine questions naturally. While prompt engineering optimizes responses, exploration and serendipitous discovery often emerge from unstructured queries, making rigidly effective prompts less beneficial for learning.

LLMs foster curiosity and self-directed learning in several ways. They personalize responses by analyzing a child's queries and suggesting targeted follow-ups. They also provide a low-stakes environment where children feel comfortable experimenting with questions, enhancing their inquiry-based learning\cite{abdelghani_guiding_2024}. Rather than just providing answers, LLMs offer hints that encourage deeper questioning, following best practices in discovery learning \cite{Holmes2019}. Additionally, their conversational style promotes engagement, acting as patient, non-judgmental companions that support gradual exploration and confidence-building in learning.

\subsection{Focussing LLMs}
Retrieval augmented generation (RAG) combines the generative power of LLMs with external knowledge retrieval systems to produce more accurate and contextually informed outputs\cite{Lewis2020}. This augmented approach addresses key limitations of stand-alone generative models, such as hallucination and factual inaccuracy, by incorporating relevant information retrieved from structured databases, corpora, or the web during the generation process.  The process typically involves two main steps.  The first is to identify and retrieve relevant documents or information from alternative sources. The second  passes these texts into the generative model as additional context. The model integrates this information with its internal knowledge to produce responses that more strongly influenced by the new knowledge.

By integrating external information sources with its inherent generative abilities, RAG can substantially reduce the risk of generating misleading information. The retrieved context acts as a dynamic, evidence-based grounding layer, enhancing the reliability of the outputs produced by the generative model.  Incorporating verifiable information from external sources helps mitigate the phenomenon of hallucination that is common in large language models. The retrieved content can also provide richer context for generating nuanced responses.

Retrieval augmented generation has found applications across a range of domains: conversational AI, in enhancing chatbot responses by grounding them in current events or factual databases; question answering - improving the precision of responses in systems that need to answer fact-based queries; content creation, by supporting writers and researchers by providing dynamically retrieved background information that complements the creative process. It is the ability to focus the LLMs onto specific areas that presents an interesting opportunity for educationalists and technology designers alike.

\section{Self-ethnographic experiments and observations}
This research is anecdotal and not scientifically rigorous, but it offers food for thought. My children are revising for their GCSE exams, the country-wide qualification for 16 year olds in the UK. I can help with Maths, but History—especially American and Medical History—is a challenge for me. Not only don't I know the topics, I don't know how they are supposed to answer the questions in the exam --- what structures for answers are effective and what are not (and I'm not alone \cite{sentance_parental_2020}. I can certainly use Google to find specific answers to factual questions, but that's not what they need. What they need is their teacher.

To assist them, we discussed what support they needed.  They explained that the different questions had different styles - the first would be comparing two sources, the second asking for options on a contemporaneous report, and that there were advised strategies and structures applicable to each. They also noted that much other online material was not done in the same way that their exam board wanted things, and so seems much less relevant. I then created a RAG-based LLM, tailored to their syllabus and exam board. I fed it the syllabus, study guides, notes, past papers, and online content, then crafted specific prompts to guide interaction, so that we could, for example, ask it for a topic in the style of Question 1, or Question 4, for example, and get material just from the syllabus being questioned in a familiar way.  Their use of the system was informally observed and discussed over a period of 6 weeks from early February to March 2025. It was, of course, not the success I had hoped: I envisaged them spending time late into the night chatting to it,  exploring history like never before.  They didn't.  But it became a valuable tool. We use it to generate practice questions, submit answers, and receive targeted feedback based on their exam format. It explains what’s missing, offers model answers, and adapts to their syllabus. Because it’s non-judgmental and patient, they feel more confident using it. Their learning is active and engaging, making it preferable to passive revision.  As a result, I’ve stepped back from helping with History, and they rely less on school revision sessions. The LLM offers more tailored, immediate, and relevant support than a teacher can in a group setting. This suggests a significant shift in education: LLMs can deliver personalized guidance and feedback at scale. Teachers roles are changing, and their focus may shift to higher-order skills—critical thinking, source evaluation, and problem solving.

\section{Interaction design in a changed world}
If I have convinced you that significant change is on the way within classrooms and possibly even to the delivery and nature of education, the question for our community is how this might impact us.  I think there are five significant influences to take into account, each discussed below.

\subsection{A shift from scroll, point, and click interfaces to conversational ones}
The interaction model for LLMs, and their increasing use cases, are conversational in style.  (By conversational, I am including both textual (the more common) and verbal communication:  voice is to be welcomed on accessibility grounds but at present tends to suffer from extraneous noise interruptions, and more obvious latency.)  We do not need to creating perfect one-shot queries to identify the answer, which makes this different from search.  Search (at least on major search engines) retains no knowledge of the previous search you preformed, and you are left to try again if the answers you want are not there.  With a LLM you can ask it to refine parts of answers, come back to earlier topics, and meander towards a resolution.  Because of the iterative, contextual nature, the exchange is much more human-like than most other forms of human-computer interaction the majority of people have experienced.  Usage models of LLMs can vary: one interaction can be short and intense, never to be revisited, whilst others may continue specific themes over many weeks or months. Even now, the capabilities of non-LLM AI systems like Alexa or Google home are quirky, but LLMs now have a very low barrier to entry and impressive skills. 

Compare this to the interaction styles common now: the WIMP system with point scroll and click is still strong, though the continuous scroll of TiKTok has become huge.  In terms of widespread use, ongoing conversational interactions are not common.  Since users like what they know, and are finding conversational interaction effective and powerful, there is likely to be a demand for such systems in areas they have not previously impacted.

\subsection{An expectation of continued context awareness across multiple interactions}
As children become used to models that know specifics about their circumstances (which exam board they are using, for example), they will begin to expect to receive personalised responses. Add to this that LLMs can retain context across conversations, knowing what your interests are, and can use this to shape their responses to to decide which branch of an exchange to take, children find themselves interacting with systems that know about them personally.  This is likely to lead to expectations for other interactive systems to 'know' wheat they did before to do it again. or to understand that they've already expressed a preference for, say, rock music and so won't appreciate a suggestion of Rick Astley\cite{astley_notitle_1987}.

This is happening in many areas of professional practice: many people now code in English. We can simply throw the problem at the LLM in a few sentences and it can produce code to resolve it.  It rarely works first time on all cases, but these new systems allow us to define aspects of problems, refine code segments, refactor things, and even change languages.  We build up a picture of a problem over time and iterate code towards solutions.  This works very well --- it is colloquially known as vibe coding, a term create by Kaparthy, a co-Founder of OpenAI \cite{karpathy_theres_2025}.  The focus is not on getting it right, at least not immediately.  Instead it's on exploring where initial ideas take you and seeing what works and what doesn't, and following promising routes and improving them (how far away from an exam have we now come?!).  Because this vibe approach is becoming more common everywhere, users will naturally have a much higher expectation of iterative refinement across all forms of interaction. 

\subsection{A lack of tolerance for highly specific systems}
Since LLMs can provide responses to anything, and context-switch in an instant. children are likely to become used to exploring numerous different things, perhaps in parallel, and may find the specific focussed nature of other interactions limiting.  Sometimes it is important be be focussed on one thing, and be constrained by technology to do just that, but if our experiences have been much more relaxed an fluid, we are less likely to be impressed by a one-trick system.  How this is likely to influence designs is hard to ascertain --- not everything should be or needs an LLM, but it seems likely that APIs to allow data export and integration with other systems will become much more common and we will produce software components rather than stand-alone software systems.  This generality changes the perceptions and nature of devices.

\subsection{Explainable systems and trust}
Explainable systems (explainable AI or xAI) are able to peer inside the black box of complex machine learning systems (like LLMs) in order to understand why the results are as they are.  One of its benefits may be to allow us to question results, to detect hallucinations, and perhaps to identify fake news or disinformation.  Only by interrogating a system and having it explain its responses can we get clarity on its response, and we need this in order to build trust in the system,  understand its capabilities and, more importantly, its limitations.

It is, perhaps, more of a hope than an expectation, but systems will in future need to be able to justify themselves to their users in order for users to trust them --- and if this is true in one context then it is likely to exist in others. and so we may have to ensure that the workings of our designs are transparent enough such that trust can be formed.  This would be a good outcome: savvy users ensuring they believe what the machines tell them before acting on it.  However, early signs are less positive: the impressive initial capabilities of the systems are so strong that we tend to abrogate responsibility and agency because it seems to know a lot abut everything and can do lots of things we don't like doing.  If it can summarise a meeting so quickly, we are tempted to think, it must be great at everything else too.  And also, human nature being what it is, if a system can provide an answer that appears immediately acceptable, we often prefer to take it at face value and move on to something else, rather than spend the time to check it.

\subsection{LLM driven design}
Children coming through the education system now are the designers of tomorrow.  We are also able to adapt and embrace these systems. Zhou et al. \cite{zhou_exploring_2024} surveyed LLM usage across practising UX designers, showing ChatGPT supporting UX design though offering design guidelines, constructing user profiles, and simulating stakeholders. They note the challenges are around understanding complex design problems and prototyping design ideas:  the essence of creative problem solving, sketching ideas and combining concepts, is still a more human preserve.  However, they represent a formidable additional tool to our palette --- for example, Liang et al.\cite{liang_automated_2018} discuss how generative AI can be used to create storyboards.  It is therefore possible that future designers can be more informed, better able to integrate new ideas, find it easier to create communicative concepts, and are freed to concentrate on idea generation and problem solving. 

\section{Conclusion}
The impact of LLMs is only just starting to be felt in the design community. The biggest shift is likely to be in expectations and demands from users, requiring us to evolve our perspectives, default interaction models, and approach to design. It is only fitting that we let ChatGPT (4.0) have the last word on this.  I asked:\\
\emph{"summarise areas of impact on the design community for children that this paper has missed"}\\
and then had it summarise its rather comprehensive response:\\
\emph{"Key overlooked impacts on child-focused design: \textbf{Social Skills} No mention of how AI may affect children's social development. \textbf{Safety} Ignores risks like bias, overuse, and harmful content.  \textbf{Collaboration} Overlooks tools for group learning.  \textbf{Inclusivity} Lacks focus on diverse user needs. \textbf{Play} Misses LLMs' role in creative activities."}\\
I agree.  But we have no more space to explore these insights here.
\bibliographystyle{ACM-Reference-Format}
\bibliography{references}
\end{document}